\documentclass[twoside,twocolumn,9pt]{article}
\usepackage{extsizes}
\usepackage[numbers,sort&compress]{natbib}
\usepackage[left=1.75cm, right=1.75cm, top=2cm, bottom=2.0cm]{geometry}
\usepackage{sectsty}
\allsectionsfont{\scshape}
\usepackage{graphicx}
\usepackage{array}
\usepackage[T1]{fontenc}
\usepackage[dvipsnames]{xcolor}
\usepackage[english]{babel}
\usepackage[hyperfootnotes=false,hidelinks]{hyperref}
\usepackage{amsmath,amssymb}
\usepackage{xspace}
\usepackage{pdfpages}

\newcommand\Rmin{\ensuremath{R_{\mathrm{min}}}\xspace}
\newcommand\Rone{\ensuremath{R_{\mathrm{1}}}\xspace}
\newcommand\Rtwo{\ensuremath{R_{\mathrm{2}}}\xspace}
\newcommand\Fpeak{\ensuremath{F_{\mathrm{peak}}}\xspace}
\newcommand\Speak{\ensuremath{S_{\mathrm{peak}}}\xspace}
\newcommand\Srup{\ensuremath{S_{\mathrm{rup}}}\xspace}
\newcommand\Scn{\ensuremath{S_{\mathrm{rup, Newt}}}\xspace}
\newcommand\Scv{\ensuremath{S_{\mathrm{rup, elast}}}\xspace}
\newcommand\epsv{\ensuremath{\dot{\varepsilon}_{\mathrm{v}}}\xspace}
\newcommand\epsc{\ensuremath{\dot{\varepsilon}_{\mathrm{c}}}\xspace}
\newcommand\Vs{\ensuremath{V_{\mathrm{s}}}\xspace}

\newcommand\tc{\ensuremath{t_{\mathrm{c}}}\xspace}
\newcommand\Sc{\ensuremath{S_{\mathrm{c}}}\xspace}
\newcommand\lr{\ensuremath{\lambda_{\mathrm{R}}}\xspace}
\newcommand\lz{\ensuremath{\lambda_{\mathrm{e}}}\xspace}
\newcommand\tauz{\ensuremath{\tau_{\mathrm{z}}}\xspace}

\begin{document}

\twocolumn[
  \begin{@twocolumnfalse}
\vspace{2cm}
\begin{center}
    \noindent\Huge{\textbf{Axial forces in capillary liquid bridges of polymer solutions$^\dag$}} \\
    \vspace{1cm}
    \noindent\large{Sreeram Rajesh,$^{a}$ Riley S. Tinianov,$^{a}$ Jooyeon Park,$^{b}$ and Alban Sauret$^{\ast b, c}$} \\
    \vspace{1cm}
    \textbf{\textsc{Abstract}}
    \vspace{2mm}
\end{center}

\noindent\normalsize{Liquid bridges form between particles during wet mixing with binders or by condensation due to ambient humidity. The consequences of capillary bridges can be quite drastic, creating macroscopic cohesion, as seen in sandcastles and in the formation of particulate agglomerates. Bulk effects in cohesive particles arise from forces generated by capillary bridges, so particle-scale measurements are needed to develop predictive models. Most existing studies at the particle scale assume Newtonian liquids. Yet many binders in industry and in the environment can exhibit viscoelastic behavior. In this study, we measure the axial force generated by liquid bridges of viscoelastic polymer solutions between two spherical beads during controlled uniaxial separation. We vary the polymer concentration, separation velocity, and particle size, and track the force as the bridge thins and ruptures. At quasi-static rates, the axial force remains dominated by capillarity and is not significantly affected by polymer rheology. However, increasing the stretching rate increases the peak force through viscous dissipation and promotes the formation of a viscoelastic filament, thereby delaying rupture. The peak axial forces collapse when rescaled by a capillary number and particle size, while the effective rupture distance collapses with a Weissenberg number. These results provide a simple first-order particle-scale force law for polymeric binders.}

 \end{@twocolumnfalse} \vspace{0.6cm}

  ]

\makeatletter
\renewcommand*{\@makefnmark}{}
\footnotetext{\textit{$^{a}$~Department of Mechanical Engineering, University of California, Santa Barbara, CA 93106, USA}}
\footnotetext{\textit{$^{b}$~Department of Mechanical Engineering, University of Maryland, College Park, MD 20742, USA}}
\footnotetext{\textit{$^{c}$~Department of Chemical and Biomolecular Engineering, University of Maryland, College Park, Maryland 20742, USA. E-mail: asauret@umd.edu}}
\footnotetext{\dag~Electronic Supplementary Information (ESI) available. See DOI: 10.1039/cXsm00000x/}
\makeatother

\section{Introduction}
\label{sec:Intro}


\begin{figure*}[htpb]
\centering
  \includegraphics[width=0.98\textwidth]{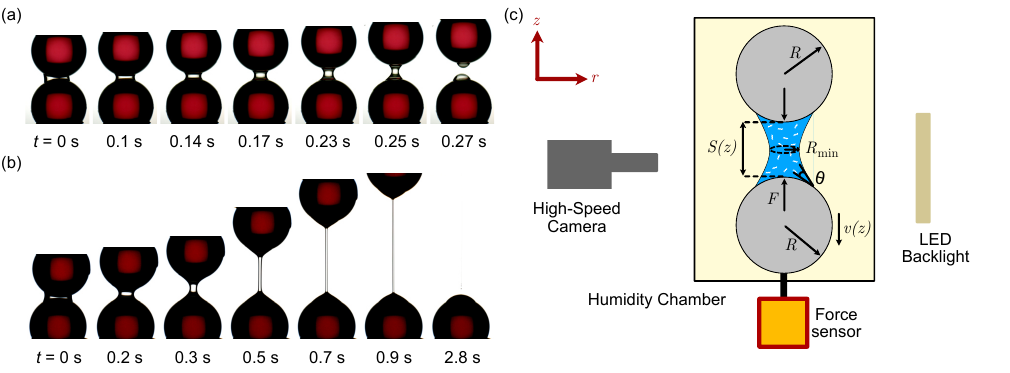}
  \caption{Thinning of a liquid bridge of (a) Water, and (b) 4M PEO with 1\% by mass concentration prepared in water. The liquid bridge is positioned between two beads of radius $R = 2\,\mathrm{mm}$ and separated with a velocity $v = 10\,\mathrm{mm/s}$. (c) Schematic of the experimental setup.}
  \label{fig:setup}
\end{figure*}

Liquid bridges between solid particles are ubiquitous, appearing in contexts ranging from soil moisture~\cite{haines1925studies, carminati2017liquid, benard2021physics} to industrial granules and wet agglomeration processes~\cite{reynolds2022extensional, calabrese2026emergent, saingier2017accretion, jop2021wet}. Many natural and industrial solids are heterogeneous, for instance, due to extracellular polymeric substances (EPS) in moisture trapped between soil particles~\cite{malarkey2015pervasive} or the use of polymeric binders in industrial slurries~\cite{reynolds2022extensional,lee2024extensional,zhang2021polymer,bitsch_novel_2014}. In high-intensity erosion and debris flows, rapid deformations with large localized strain rates at particle scales can trigger microscale polymeric cohesion~\cite{kameda2021influence, kostynick2022rheology, orts2007use, zeinali2025using}. Similarly, precise modeling of particle-scale interactions with binders is critical for industrial granulation, fluidization, and asphalt pavement design~\cite{ennis1991microlevel, zhou2023wetting, beainy2013viscoelastic}. However, a direct measurement of polymeric axial forces between a pair of individual particles is lacking in the literature.

\smallskip

Particle-scale axial forces in Newtonian liquid bridges, shown in Fig.~\ref{fig:setup}(a), are well-established~\cite{mason1965liquid, pitois_liquid_2000, willett_capillary_2000}. Early investigations focused on theoretical descriptions of capillary cohesion in soil~\cite{haines1925studies, fisher1926capillary} and submerged, gravity-free measurements of inter-particle adhesion~\cite{mason1965liquid}. Lian \textit{et al.}~\cite{lian1993theoretical} later provided numerical solutions for capillary force and rupture distance, which were validated experimentally using silicone oil by Willet \textit{et al.}~\cite{willett_capillary_2000} and Pitois \textit{et al.}~\cite{pitois_liquid_2000}. Pitois \textit{et al.} and Ennis \textit{et al.}~\cite{ennis_1990_influence} extended these measurements to fully wetting viscous fluids. Force measurements in industrially relevant water-based solutions add challenges involving evaporation, contact angle hysteresis~\cite{shi_dynamic_2018}, gravitational effects~\cite{mazzone_effect_1986, wang_mathematical_2022, mielniczuk_characterisation_2018}, and three-body interactions~\cite{wang_capillary_2017}. More recently, more complex expressions have been developed for capillary bridges between spheres, including models for perfectly and partially wetting particles~\cite{kruyt_analytical_2017,lian_closedform_2016,zhao_nonperfectly_2018,zhao_suction_2020}. The present work extends these investigations to include the impact of polymeric viscoelasticity. Indeed, although some studies have considered the breakup of viscoelastic liquid-bridge and the liquid transfer between separating solid surfaces with free or pinned contact lines, for instance, in geometries relevant to printing applications, direct measurements of the axial force induced by a polymeric bridge between two spherical particles remain elusive~\cite{chen2021viscoelastic,pingulkar2021liquid,sankaran2012effect,lee2013computational}.

\smallskip

Inter-particle axial forces are primarily controlled by capillarity. The capillary thinning dynamics of Newtonian and non-Newtonian liquid bridges are well established~\cite{entov_effect_1997,  bazilevskii_dynamics_2015, mckinley_f_2002, anna_interlaboratory_2001, rodd_capillary_2005, mckinley2000extract, rajesh_transition_2022, dekker_when_2022, dinic_pinch-off_2017, gaillard_when_2024, gaillard_beware_2024}. For Newtonian liquid bridges, the regime in which the breakup occurs (viscous, inertial, and viscous-inertial) is controlled by the Ohnesorge number, so the relevant pre-elastic thinning dynamics depend on both the liquid properties and the bridge size~\cite{li_sprittles_2016}. Previous investigations have shown that polymers modify capillary flow in uniaxial extension, where the thinning dynamics are demarcated by distinct Newtonian and viscoelastic regimes, introducing new length scales to the flow~\cite{entov_influence_1984, Amarouchene2001, Tirtaatmadja2006, rajesh_transition_2022, calabrese_how_2024}. Related studies on viscoelastic particulate suspensions further showed that coupling polymer elasticity with microstructural heterogeneity modifies the transition to the viscoelastic regime~\cite{thievenaz2021viscoelastic}. Techniques such as FISER~\cite{mckinley_f_2002, anna_interlaboratory_2001}, CaBER~\cite{rodd_capillary_2005, mckinley2000extract}, droplet pinch-off~\cite{rajesh_transition_2022, rajesh_pinch-off_2022, dekker_when_2022}, Dripping-onto-Substrate (DoS)~\cite{dinic_pinch-off_2017, dinic2019macromolecular} and the Slow Retraction Method (SRM)~\cite{gaillard_when_2024, gaillard_beware_2024} have established that polymeric axial flows exhibit an initial Newtonian regime, followed by a sharp coil-stretch transition to a viscoelastic regime characterized by cylindrical ligaments. We show this for a capillary liquid bridge between spherical particles in Fig.~\ref{fig:setup}(b). Recent studies indicate a smoother transition at semi-dilute entangled concentrations due to lower critical coil-stretch strain rates~\cite{gaillard_when_2024, rajesh_transition_2022}. More generally, recent work has shown that the onset of the elasto-capillary regime is not determined solely by a single relaxation time, but also depends on the initial stretching history, the bridge geometry, and finite extensibility of the polymers~\cite{clasen_dilute_2006,campo_deano_srm_2010,gaillard_beware_2024,gaillard_onset_2025,aisling_initial_rate_2024}. In the Newtonian regime, coiled polymers contribute to viscous dissipation that scales with concentration, generating larger axial forces than the solvent. Consequently, the combined effects of increased force and delayed rupture distinguish polymeric bridges from Newtonian ones, requiring modified bulk descriptions for granular-polymer mixtures.

\smallskip

This study investigates the axial forces in polymeric liquid bridges across a range of polymer concentrations and separation velocities, $v$, spanning four orders of magnitude. We specifically use a constant separation velocity, contrasting with the exponential or gravity-driven profiles in previous studies~\cite{mckinley2000extract, rajesh_transition_2022}, due to the relevance of grain kinematics in Discrete Element Method (DEM) models. Experimental methods and polymer rheology characterization are detailed in \S~2. Section~3 examines these forces under quasi-static conditions ($v = 0.01~\mathrm{mm/s}$), where thinning is strictly controlled by capillarity, as well as under dynamic separations, where viscoelastic effects become prominent. We then present a first-order analytical framework to describe the evolution of the force across both regimes. In \S~4, we develop dimensionless scaling laws: we collapse the peak force---which governs the bridge's adhesive strength---using the Capillary number, confirm its linear scaling with particle size, and rescale the extended rupture distance using the Weissenberg number. Finally, \S~5 provides our concluding remarks.

\medskip

\section{Experimental Methods}
\label{sec:exp_methods}
\subsection{Setup}

The custom-built experimental setup [Fig.~\ref{fig:setup}(c)] features a Futek LSB 200 force sensor with an accuracy of $\pm 5~\mathrm{\mu N}$. Data acquisition is performed at $10$--$100~\mathrm{Hz}$. This rate is well below the sensor's $140~\mathrm{Hz}$ resonant frequency, yet sufficient to resolve the viscoelastic dynamics governed by the polymer relaxation time ($\approx 100~\mathrm{ms}$). To minimize mechanical noise, the lower bead is mounted on the force sensor, which rests on an active vibration-isolation system (Nexus, Thorlabs). Meanwhile, the upper bead is attached to a linear translation stage (Thorlabs NRT150) that provides $0.01~\mathrm{mm}$ positional precision and a velocity resolution of $0.01~\mathrm{mm/s}$. The stage's maximum travel ($\sim 30~\mathrm{mm}$) exceeds the maximum rupture distance observed in this study (see \S~\ref{subsec:rupture_distance}). Both beads (ruby-doped sapphire, Edmund Optics) have a radius $R = 2~\mathrm{mm}$ and a surface roughness on the order of a few microns. Additionally, a subset of experiments utilizes particles with radii ranging from $R = 1$ to $3~\mathrm{mm}$. A custom environmental chamber encloses the particle pair to maintain a relative humidity $> 80\%$, effectively suppressing evaporation over the experimental timescale~\cite{huisman2023evaporation}. Finally, a high-speed camera (Phantom VEO 710) records the liquid bridge dynamics during separation at $100$--$1000~\mathrm{FPS}$. A validation test for the setup with AP100 silicone oil is described in \S~S1 in the Supplementary Material.
\medskip

\subsection{Methodology}
\label{subsec:methodology}

The force measurement procedure is as follows: a solution droplet ($V = 0.5$--$1~\mu\mathrm{L}$) is deposited onto the lower bead using a micropipette ($0.1$--$2.5~\mu\mathrm{L}$, Eppendorf). Since polymer viscosity and setup geometry make precise dispensing challenging, the actual bridge volume is measured using a custom image-processing routine (see Supplementary Material \S~S5). The upper bead is then lowered onto the droplet and oscillated vertically by a few micrometers at low velocity to ensure axial symmetry without disrupting the initial polymer microstructure. The inter-particle gap is subsequently zeroed by monitoring force readings, which become negative upon contact. Following this, data acquisition begins by synchronizing the sensor, camera, and translation stage to track the force $F$ as a function of the gap $S$. Separation velocity is varied within the range $v \in [0.01, 10]~\mathrm{mm/s}$. Finally, force data and high-speed videos are post-processed using custom Python and ImageJ routines.

\medskip


\begin{figure*}[!t]
\centering
\includegraphics[width=0.98\textwidth]{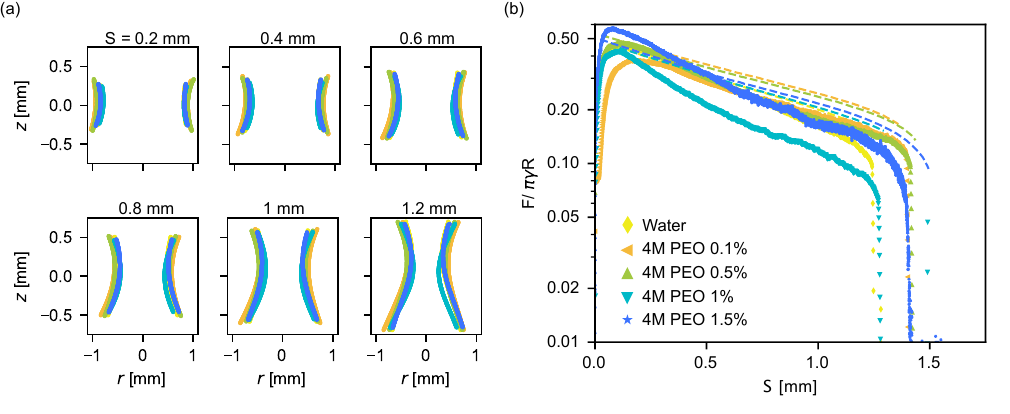}
\caption{(a) Contours of a quasi-static liquid bridge of $c = 0$--$1.5$\% 4M PEO in water for $S = 0.2$ to $S = 1.2\,\mathrm{mm}$ at separation velocity $v = 0.01\,\mathrm{mm/s}$ for various gaps $S$ between the particles. (b) Quasi-static axial force as a function of the gap $S$. The dashed lines show the prediction of the reduced capillary model in Eqn.~\ref{eqn:capillary_force}.}
\label{fig:quasistatic_dynamics}
\end{figure*}

\subsection{Rheology of the Polymers}
\label{subsec:rheology}

Polymer solutions are prepared by dissolving polyethylene oxide (PEO, Sigma-Aldrich) powder with molecular weight $M_{\rm w} = 4000~\mathrm{kg/mol}$ in deionized (DI) water; these are hereafter referred to as 4M PEO. Mixtures are homogenized on a roller (ThermoFisher Scientific\textsuperscript{TM}) at $40~\mathrm{RPM}$ for 24 hours prior to use. The intrinsic viscosity is $[\eta] = 0.072{M_{\rm w}^{0.65}} = 1.41~\mathrm{m^3/kg}$, yielding an overlap concentration $c^{*} = 0.77/[\eta] = 0.055\%$~\cite{Tirtaatmadja2006}. We test concentrations $c = 0, 0.1, 0.5, 1,$ and $1.5\%$, spanning the semi-dilute to dense regimes ($ 2 \leq c/c^* \leq 30$). To minimize chain degradation, solutions of each concentration are prepared individually rather than by diluting a stock. The shear rheology for the solutions, summarized in Fig.~S2(a) in Supplementary Material, exhibits strong shear-thinning behavior fitted by the Carreau-Yasuda model. Fig.~S2(b) details the corresponding viscous and elastic moduli.

\smallskip

The extensional rheology is summarized in Fig.~S3. The relaxation time $\lr$, determined via droplet pinch-off, scales as $c^{0.7}$, consistent with previous observations~\cite{Tirtaatmadja2006, rajesh_transition_2022}. The measurement protocol is outlined in \S~S3, and further details are available in the literature~\cite{rajesh_transition_2022, dinic_pinch-off_2017, Amarouchene2001, Tirtaatmadja2006}. We use this relaxation time, $\lr$, as a reference timescale for analyzing the axial force and rupture dynamics of the liquid bridge. However, recent studies have emphasized that relaxation times inferred from capillary-thinning experiments can depend on the deformation history, bridge size, finite extensibility, and the molecular-weight distribution sampled during stretching; consequently, such thinning timescales should be interpreted operationally rather than as unique geometry-independent material constants~\cite{clasen_dilute_2006,campo_deano_srm_2010,gaillard_beware_2024,gaillard_onset_2025,aisling_initial_rate_2024,calabrese_polydispersity_2025}.

\medskip

\subsection{Interfacial Properties}
\label{subsec:interface_properties}

Interfacial properties are summarized in Supplementary Material Fig.~S4. Fig.~S4(a) shows that adding $0.1\%$ polymer reduces the surface tension from $\gamma = 72.8~\mathrm{mN/m}$ for water to $\approx 63~\mathrm{mN/m}$. Subsequent concentration increases yield negligible changes, as measured by an Attension Tensiometer. While direct measurement at $c = 1.5\%$ proved difficult, we assume $\gamma = 60~\mathrm{mN/m}$ consistent with high-molecular-weight solutions~\cite{wonakim1994molecular}. The solid-liquid contact angle $\theta$ in Fig.~S4(b) is extracted via custom Python image processing as a function of separation distance $S$. At small $S$, $\theta$ is initially large because the bridge remains convex at very small gaps, and it then decreases as the meniscus evolves toward a concave shape with increasing separation, before stabilizing at larger gaps~\cite{xiao_capillary_2020}. Although $\theta$ varies slightly with concentration and volume, it stabilizes beyond initial gaps to $\theta \in [30^{\circ}, 50^{\circ}]$ for all $c$.

\medskip

\section{Results}

\subsection{Quasi-static Axial Forces}
\label{subsec:quasistatic}

In this subsection, we present axial forces measured under quasi-static conditions. For the bridge volumes used here, a truly static configuration is difficult to maintain because the liquid slowly drains under gravity. We therefore use the lowest accessible separation velocity, $v = 0.01~\mathrm{mm/s}$, as a practical quasi-static limit. Figure~\ref{fig:quasistatic_dynamics}(a) shows that, at a given gap, the liquid-bridge contours remain very similar across 4M PEO solutions with $c = 0$--$1.5\%$. Consistently, the rescaled force profiles in Fig.~\ref{fig:quasistatic_dynamics}(b) also vary only weakly with concentration. This indicates that the quasi-static response is controlled mainly by capillarity and bridge geometry, while polymer rheology plays only a minor role in this regime. This interpretation is consistent with the interfacial measurements reported in Fig.~S4. Under these conditions, the axial force can therefore be estimated from the liquid-bridge shape, as for water~\cite{kruyt_analytical_2017}. Although the measured liquid bridge volume varies slightly across concentrations ($V = 0.6$--$0.9~\mu\mathrm{L}$), it does not significantly influence the observed curvature. As a result, capillarity dominates, and we expect similar axial forces for $c = 0$--$1.5\%$. Therefore, under quasi-static conditions, capillary forces can be estimated by quantifying the liquid-bridge shape, since the primary difference between solutions is surface tension, as in water.

\smallskip

Deriving an exact analytical solution for the capillary force between particles requires solving the Young-Laplace equation, which is generally intractable~\cite{orr1975pendular}. Therefore, classic work relied on numerical solutions or fitted expressions based on numerical integration of the Young-Laplace equation~\cite{lian1993theoretical,willett_capillary_2000,pitois_liquid_2000}. More recently, analytical or near-closed-form expressions have been developed for perfectly wetting bridges, finite contact angles, suction-controlled bridges, and unequal contact angles between the two particles~\cite{kruyt_analytical_2017,lian_closedform_2016,zhao_nonperfectly_2018,zhao_suction_2020,wu_unequal_contact_angle_2020}. In the present work, however, we seek a reduced description that can be used consistently in both the quasi-static regime and the later dynamic viscoelastic regime. We therefore express the force in terms of the measured minimum neck radius of the liquid meniscus, $\Rmin$, which is directly accessible from the bridge images throughout stretching. Such neck-based descriptions have been widely used for extensional flows of inviscid, viscous, and viscoelastic liquids~\cite{mckinley2000extract,Eggers1997,mckinley_f_2002}, and provide a convenient framework between the static force problem and the thinning dynamics.

\smallskip

To describe the capillarity-dominated axial force on the lower bead, we consider the liquid volume $V_s$ below the minimum meniscus radius $\Rmin$, as shown in Fig.~\ref{fig:setup}(c):
\begin{equation}
F = 2\pi\gamma\Rmin - \pi \Rmin^2 \Delta P - \Vs \rho g
\label{eqn:capillary_force_willet}
\end{equation}
The first term represents the axial contribution of surface tension along the perimeter $2\pi\Rmin$. The second term accounts for the Laplace pressure difference across the curved interface, while the third term corrects for gravitational distortions~\cite{willett_capillary_2000}. The pressure difference $\Delta P$ is given by the Young-Laplace equation:
\begin{equation}
\Delta P = \gamma \left[ \frac{1}{\Rone} + \frac{1}{\Rtwo} \right]
\end{equation}
where $\Rone$ and $\Rtwo$ are the principal radii of curvature of the liquid-air interface. We approximate the first principal radius (inside the liquid) as $\Rone \approx \Rmin$. Image analysis indicates that the local curvature near the neck is dominated by the azimuthal component, such that $1/\Rone \gg 1/\Rtwo$. This assumption is also consistent with the contact-angle measurements for these water-based solutions [see Fig.~S4(b)]. Finally, for the reduced description adopted here, we neglect the gravitational term. These approximations simplify Eqn.~\ref{eqn:capillary_force_willet} to:
\begin{equation}
F \simeq \pi\gamma\Rmin
\label{eqn:capillary_force}
\end{equation}

We emphasize that Eqn.~\ref{eqn:capillary_force} is a reduced approximation rather than a general replacement for the closed-form capillary-bridge expressions cited above. Its role here is to capture the measured force decay with a single geometric quantity that can also be tracked during dynamic thinning. Equation~(\ref{eqn:capillary_force}) is plotted against the experimental data as dashed lines in Fig.~\ref{fig:quasistatic_dynamics}(b). Even without accounting for gravitational distortions, it describes the experimental measurements of $F$ reasonably well as $ F$ decays from its maximum, eliminating the need for rigorous modeling of the liquid-bridge geometry.

\smallskip


\begin{figure*}[htpb]
\centering
  \includegraphics[width = 0.98\textwidth]{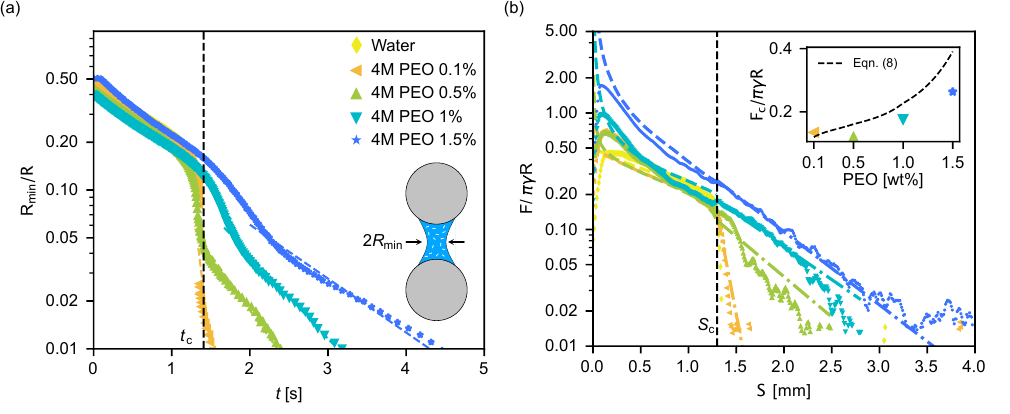}
  \caption{(a) Time evolution of the rescaled minimum neck radius $\Rmin(t)/R$ for liquid bridges of 4M PEO in water ($c = 0-1.5\%$) separating at $v = 1\,\mathrm{mm/s}$. The thinning dynamics transition from Newtonian ($t < \tc$) to viscoelastic ($t \ge \tc$). (b) Corresponding rescaled axial force $F/\pi\gamma R$ versus separation gap $S = vt$. Dashed lines correspond to the model predictions in the Newtonian (Eqn.~\ref{eqn:force_Newtonian}) and viscoelastic (Eqn.~\ref{eqn:Force_Viscoelastic}) regimes. The inset compares the fitted critical force $F_{\mathrm{c}}$ from Eqn.~\ref{eqn:Force_Viscoelastic} with the prediction from Eqn.~\ref{eqn:Critical_force_transition}.} 
  \label{fig:thinning}
\end{figure*}

Experimentally, we observe that the force initially increases to a peak value before decreasing as $S$ increases~\cite{xiao_capillary_2020,willett_hysteresis_2003,wang_capillary_2017,mielniczuk_characterisation_2018,megias2009capillary}. This non-monotonic behavior is consistent with the evolution of the liquid bridge from a convex meniscus at small gaps to a concave meniscus at larger gaps, together with the associated evolution of the contact line and contact angle~\cite{xiao_capillary_2020}. At small gaps, $\theta \gtrsim 90^{\circ}$ can produce a repulsive contribution associated with the convex meniscus. As the bridge is stretched, the attractive force reaches a maximum and then decreases as the bridge thins further~\cite{xiao_capillary_2020}. In addition, wetting hysteresis may pin the three-phase contact line and generate metastable bridge states, thereby affecting the exact position and magnitude of the force maximum. The wetting history and instantaneous bridge geometry may therefore contribute to the observed variations in the peak axial force~\cite{willett_hysteresis_2003}. The peak axial force is $F \simeq 0.18 \pm 0.03~\mathrm{mN}$ across all concentrations ($c=0$--$1.5\%$). Furthermore, the $F(S)$ profile is similar for all solutions. Minor variations, particularly for $c = 1\%$, result from a smaller bridge volume ($V = 0.6~\mu\mathrm{L}$ versus $0.7$--$0.9~\mu\mathrm{L}$ for the other concentrations). A smaller volume leads to faster force decay, as evidenced by the rupture distance scaling $S \propto V^{1/3}$~\cite{lian1993theoretical}. Consequently, the measured forces are essentially independent of $c$. For $c \geq 0.1\%$, weak viscoelastic effects appear only milliseconds before rupture and are negligible.


\subsection{Dynamic Liquid Bridges: Axial force in the pre-elastic viscocapillary regime}
\label{subsec:dynamic}

When the separation velocity $v > 0.01~\mathrm{mm/s}$, the viscoelastic regime becomes increasingly relevant. The transition between regimes is triggered by the imposed strain rate, which, when sufficiently large, leads to the onset of viscoelasticity~\cite{rajesh_transition_2022}. In velocity-controlled setups, such as FISER or the present study, we use the global kinematic measure $\varepsilon_{\rm v} = S/R$, resulting in an imposed strain rate $\dot{\varepsilon_{\rm v}} = v/R$. If $\dot{\varepsilon_{\rm v}}$ is sufficiently large, it triggers the abrupt uncoiling and extension of polymer chains~\cite{DeGennes1974}, resulting in a viscoelastic regime characterized by cylindrical ligaments.

\smallskip

Figure~\ref{fig:thinning}(a) shows the time evolution of the rescaled minimum neck radius, $\Rmin(t)/R$, for 4M PEO solutions at $v = 1~\mathrm{mm/s}$. As expected, the initial thinning regime ($t < \tc$) is dominated by Newtonian effects, resulting in similar profiles for all concentrations ($c = 0$--$1.5\%$)~\cite{rajesh_transition_2022, rodd_capillary_2005}. At $t \sim t_{\mathrm{c}}$, the dynamics become viscoelastic [dashed lines in Fig.~\ref{fig:thinning}(a)]. Although $t_{\mathrm{c}}$ varies with concentration, we indicate a representative transition time $t_{\mathrm{c}} \simeq 1.3~\mathrm{s}$ for visual guidance. For $t > t_{\mathrm{c}}$, the bridge undergoes exponential thinning:
\begin{equation}
    \Rmin \propto R_{\rm c} e^{-t/3\lz}
\label{eqn:exp_thinning}
\end{equation}
where $R_{\rm c}$ is the critical radius at the onset of the viscoelastic regime and $\lz$ a characteristic timescale for relaxation of the stretched polymers in the liquid bridge. Unlike simplified interpretations of CaBER or droplet pinch-off in which the exponential thinning rate directly returns a unique longest relaxation time, recent work shows that the apparent elasto-capillary timescale can depend on the pre-stretch history, bridge size, and finite extensibility of the polymers, and can also vary with concentration because different portions of a polydisperse molecular-weight distribution are recruited during stretching~\cite{campo_deano_srm_2010,gaillard_beware_2024,gaillard_onset_2025,aisling_initial_rate_2024,calabrese_polydispersity_2025}. We therefore interpret $\lz$ here as an effective thinning timescale for the present constant-velocity liquid-bridge protocol, rather than as a direct measurement of a unique longest material relaxation time. The observation that $\lz < \lr$ is thus consistent with incomplete chain stretching prior to the establishment of the cylindrical filament~\cite{clasen_dilute_2006}.

\smallskip 

The corresponding axial forces are shown in Fig.~\ref{fig:thinning}(b). In the Newtonian regime, the axial force measured at $v = 1~\mathrm{mm/s}$ increases with concentration. Furthermore, an increase in concentration sustains the force for a longer duration following the transition to the viscoelastic regime. In summary, both breakup time and rupture distance increase with $v$ and $c$, distinguishing viscoelastic bridges from their Newtonian counterparts under dynamic conditions.

\smallskip

In the Newtonian regime ($t < t_{\mathrm{c}}$), polymer chains in solution remain coiled, and the hydrodynamic interactions between these coils primarily drive viscous dissipation~\cite{DeGennes1974}. Assuming shear-thinning effects are negligible in this regime, the viscous contribution to the axial force can be expressed as~\cite{nase2001discrete, pitois_liquid_2000}:
\begin{equation}
    F_{\mathrm{visc}} = \frac{3}{2}K \pi \eta R^2 \frac{v}{S}
    \label{eqn:force_viscous}
\end{equation}
We use a prefactor of $3\pi/2$ instead of $6\pi$, as the former derives from the lubrication approximation~\cite{pitois_liquid_2000}, whereas the latter is associated with Stokes drag. The critical parameter here is the increased solution viscosity $\eta$ due to the presence of polymers. As a first estimate of the viscous contribution, we use an effective shear-rate-dependent viscosity extracted from the Carreau-Yasuda fit to the shear rheology [Fig.~S2(a)], evaluated at $\dot{\gamma}_{\mathrm{eff}} \sim v/\Rmin$. We emphasize that this is an effective estimate for the pre-elastic regime, not a direct equivalence between shear and extensional viscosities. Additionally, a correction factor $K = 0.5$ accounts for the finite volume of the liquid bridge~\cite{pitois_liquid_2000}.

\smallskip

The total axial force in the Newtonian regime is the sum of capillary and viscous contributions:
\begin{equation}
    F \simeq \pi\gamma\Rmin + \frac{3}{2}K\pi \eta R^2 \frac{v}{S}
    \label{eqn:force_Newtonian}
\end{equation}
Equation~\ref{eqn:force_Newtonian}, plotted as dashed lines in Fig.~\ref{fig:thinning}(b), accurately describes the experimental measurements in the Newtonian regime, except as $S \rightarrow 0$, where the modeled force diverges. As noted in \S~\ref{subsec:quasistatic}, we attribute this discrepancy to large contact angles that result in particle-particle repulsion. Altogether, we present a first-order approximation that captures the axial force for a given polymer concentration and separation velocity using independently measured material properties and the literature correction factor $K=0.5$. In the Newtonian regime, this description does not require any additional fitting to the force data.



\begin{figure}[!b]
\centering
  \includegraphics[width=0.48\textwidth]{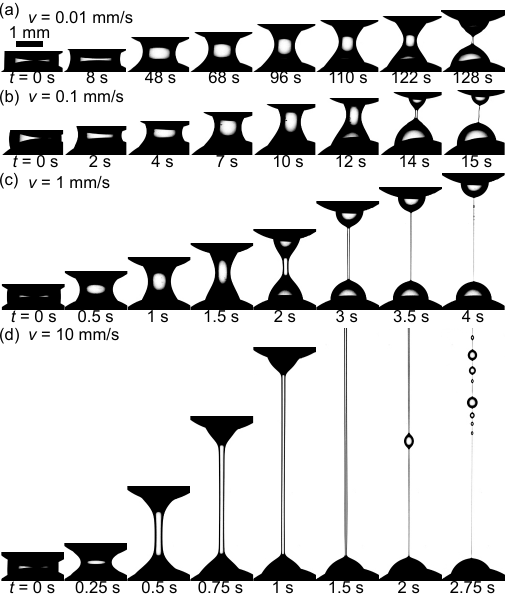}
  \caption{Temporal evolution of the liquid bridge profile for 1\% 4M PEO solution at (a) $v = 0.01\,\mathrm{mm/s}$, (b) $v = 0.1\,\mathrm{mm/s}$, (c) $v = 1\,\mathrm{mm/s}$, and (d) $v = 10\,\mathrm{mm/s}$.}
  \label{fig:force_vs_velocity_schema}
\end{figure}

\subsection{Axial Force in the viscoelastic regime}

As noted with Eqn.~\ref{eqn:exp_thinning}, the viscoelastic regime ($t \geq t_{\mathrm{c}}$) exhibits exponential thinning characterized by cylindrical ligaments. While previous studies using FISER~\cite{anna_interlaboratory_2001, mckinley_f_2002}, CaBER~\cite{van_aeken_analysis_2022, klein2009addition} and droplet thinning~\cite{bazilevskii_dynamics_2015} have estimated axial stresses in the ligaments, our focus is on the total force exerted on the particles, relevant to bulk cohesive granular flows~\cite{sharma2025experimental, richefeu2006shear}. The end geometry of the liquid bridge likely modifies this total force compared to the localized stresses in the cylindrical region.

From Fig.~\ref{fig:thinning}(b), we observe that for $t > t_{\mathrm{c}}$, the axial force decays exponentially, similar to the thinning dynamics. Motivated by the Oldroyd-B elasto-capillary thinning law and by the measured force decay, we therefore use the phenomenological form
\begin{equation}
    F = F_{\mathrm{c}} e^{-\frac{S-\Sc}{3v\lz}}
    \label{eqn:Force_Viscoelastic}
\end{equation}
where $\Sc$ is the critical gap at the onset of the viscoelastic transition, $\lz$ is the effective thinning timescale measured from the ligament evolution, and $F_{\mathrm{c}}$ is the transition force obtained from the fit. As shown in Fig.~\ref{fig:thinning}, Eqn.~\ref{eqn:Force_Viscoelastic} provides a good approximation for the axial forces in this regime. 

\smallskip

The prefactor $F_{\mathrm{c}}$ is the critical elastic force at this transition ($t = \tc$). At its onset, the critical minimum bridge radius is $R_{\mathrm{c}}$ (see Eqn.~\ref{eqn:exp_thinning})---the radius at which polymers are sufficiently stretched to dominate the dynamics. To obtain an order-of-magnitude estimate for $F_{\mathrm{c}}$, we equate the Newtonian force (Eqn.~\ref{eqn:force_Newtonian}) to the polymeric force $F \approx \pi R_{\mathrm{c}}^2 \tau_{zz}$ at $S = S_{\mathrm{c}}$, and assume that the elastic stress balances the capillary pressure ($\tau_{zz,c} \approx \gamma/R_{\mathrm{c}}$). This scaling argument gives:
\begin{equation}
    F_{\mathrm{c}} \approx \pi \gamma R_{\mathrm{c}} \left[ 1 + \frac{3K\eta R^2 v}{2\gamma R_{\mathrm{c}} S_{\mathrm{c}}} \right]
    \label{eqn:Critical_force_transition}
\end{equation}
Previous studies have shown that the critical radius scales with concentration as $R_{\mathrm{c}} \propto c^{0.15}$~\cite{rajesh_transition_2022, dekker_when_2022}. Substituting this scaling into Eqn.~\ref{eqn:Critical_force_transition} allows us to compare the model's predictions with the fitted values of $F_{\mathrm{c}}$, as shown in the inset of Fig.~\ref{fig:thinning}(b). As Rajesh \textit{et al.}~\cite{rajesh_transition_2022} noted, $R_{\mathrm{c}}$ increases more rapidly in the semi-dilute entangled regime, which explains the slight overestimation by Eqn.~\ref{eqn:Critical_force_transition}. Nevertheless, this approach establishes a framework for describing the viscoelastic axial force using parameters derived entirely from the thinning dynamics.

\smallskip

In summary, combining Eqn.~\ref{eqn:force_Newtonian} and Eqn.~\ref{eqn:Force_Viscoelastic} yields a first-order model that captures the axial forces across both the Newtonian and viscoelastic regimes in polymeric liquid bridges. The piecewise framework presented here is a simplified study---neglecting early elastic contributions at higher concentrations and approximating the elasto-capillary transition as discrete regimes. In the following sections, we further investigate these dynamic liquid bridges by varying the separation velocity $v$, thereby controlling the imposed strain rate $\dot{\varepsilon}$.

\subsection{Axial Force dependence on stretching velocities}
\label{subsec:force_vs_velocity}

To trigger the viscoelastic transition, the local meniscus strain rate $\dot{\varepsilon}$ must exceed the critical unwinding strain rate, $\epsc$, of the polymers~\cite{DeGennes1974}.  This critical rate $\epsc$ depends on polymer properties such as concentration, molecular weight, and solvent viscosity~\cite{rajesh_transition_2022, Tirtaatmadja2006}. Meanwhile, the imposed strain rate, $\epsv = v/R$, is velocity-controlled and distinct from the local strain rate $\dot{\varepsilon}$. Figure~\ref{fig:force_vs_velocity_schema} illustrates the effect of $\epsv$ on the capillary flow dynamics of a $1\%$ 4M PEO solution, where similar bridge volume is maintained across different velocities. All else being equal, we observe that the duration of the viscoelastic regime increases with $v$. In this subsection, we quantify the evolution of the axial force with the separation gap across velocities spanning four orders of magnitude ($v = 0.01 - 10\,\mathrm{mm/s}$), highlighting the maximum force, $\Fpeak$, which controls the adhesive strength of the liquid bridge in bulk granular materials~\cite{ennis_1990_influence}.


\begin{figure}[!t]
\centering
  \includegraphics[width=0.48\textwidth]{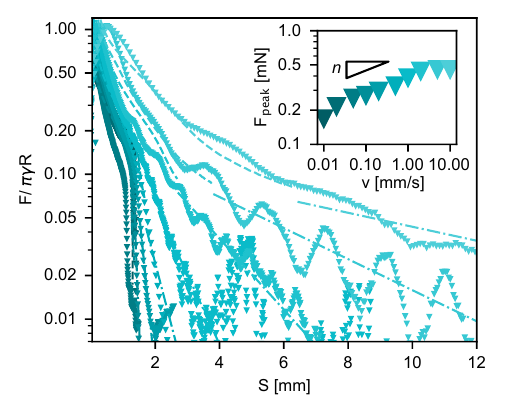}
  \caption{Axial forces in a 1\% 4M poly-ethylene oxide solution prepared in water as a function of velocity in the range $v = 0.01 - 10\,\mathrm{mm/s}$.}
  \label{fig:force_vs_velocity}
\end{figure}

\smallskip

Figure~\ref{fig:force_vs_velocity} quantifies the evolution of $F(S)$ across separation velocities $v \in [0.01, 10]~\mathrm{mm/s}$. The measured forces align well with our qualitative observations: an increase in $v$ prolongs the persistence of the force signal. In addition to the increase in viscous dissipation, rate-dependent changes in the contact angle may also contribute to the early-stage force response, since dynamic contact angle hysteresis in liquid bridges becomes more pronounced as the loading rate increases, thereby modifying the capillary-force response~\cite{shi_dynamic_2018}. Quantifying this phenomenon is highly relevant to the bulk transport and advection of cohesive grains---such as in fluidization or silo flows~\cite{sharma2026gravity, philippe2013localized} ---where large inter-particle distances are typical, and the presence of polymers can significantly modify particle dynamics. To model the force across these velocities, we apply Eqn.~\ref{eqn:force_Newtonian} in the Newtonian regime and Eqn.~\ref{eqn:Force_Viscoelastic} in the viscoelastic regime. Both equations describe $F$ accurately using parameters extracted directly from shear rheology ($\eta$) and thinning dynamics ($\Rmin$, $\lz$).

\smallskip

The inset of Fig.~\ref{fig:force_vs_velocity} shows the peak force, $\Fpeak$, required to separate the particles, which increases with velocity. This increase in $\Fpeak$, coupled with the extended duration of the thinning dynamics, is attributed to enhanced viscous dissipation caused by the uncoiling of a larger fraction of polymers at higher imposed strain rates ($\epsv$). In the following sections, we further explore the relationship between $v$ and $\Fpeak$.

\medskip

\section{Discussion}
\label{sec:discussions}

\subsection{Strength of a liquid bridge}


\begin{figure}[!b]
\centering
  \includegraphics[width=0.48\textwidth]{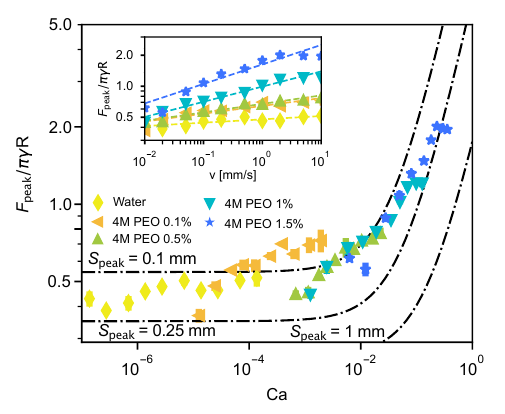}
  \caption{Bridge strength ($\Fpeak$) measured for polymer solutions across various concentrations $c = 0 - 1.5\%$ as a function of the capillary number, ${\mathrm{Ca}}$. Inset: the bridge strength dependence on separation velocities.}
  \label{fig:peak_force}
\end{figure}

Macroscopic cohesive granular flows are typically characterized by their bulk yield stress, $\tauz$, which is linked to the particle-scale bridge strength via the Rumpf scaling~\cite{rumpf1974wissenschaft}: $\tauz \propto \Fpeak/R^2$. Although macroscopic techniques such as powder shear cells are commonly used to measure static yield stress, they are often expensive and prone to high variability. Predicting $\tauz$ directly from particle-scale measurements of $\Fpeak$ provides a robust alternative and establishes a framework for developing structure-property relationships~\cite{mitarai_wet_2006, herminghaus__dynamics_2005, sharma2025experimental}. Beyond static strength measurements, particle-scale cohesion is also known to control wet aggregate growth and the flow of cohesive granular assemblies, underscoring the need for model of force laws at the particle level~\cite{saingier2017accretion, gans2023collapse, sharma2024effects}. In this section, we characterize the evolution of $\Fpeak$ in polymeric liquid bridges across varying polymer concentrations and separation velocities spanning four orders of magnitude. By rescaling this relationship with the dimensionless Capillary number, $\mathrm{Ca}=\eta v/ \gamma$, we demonstrate that the scaling framework originally developed for viscous Newtonian liquids~\cite{ennis_1990_influence} can be successfully extended to polymer solutions.

\smallskip

The inset of Fig.~\ref{fig:peak_force} shows $\Fpeak$ for 4M PEO solutions ($c = 0$--$1.5\%$) across separation velocities $v \in [0.01, 10]~\mathrm{mm/s}$. As expected, $\Fpeak$ scales with both velocity and polymer concentration. Because $\Fpeak$ is measured within the Newtonian regime, a Newtonian fluid-based description provides a good starting point, despite polymeric contributions to the viscosity. Motivated by Ennis \textit{et al.}~\cite{ennis_1990_influence}, we evaluate Eqn.~\ref{eqn:force_Newtonian} at the gap $S = \Speak$, where the peak force, $\Fpeak = F(S =\Speak)$, occurs. We then rescale this expression using the Capillary number, yielding:
\begin{equation}
    \frac{\Fpeak}{\pi\gamma R} = \frac{\Rmin}{R} + \frac{3}{2}K\mathrm{Ca}\frac{R}{\Speak}
    \label{eqn:rescaled_viscous_force_capillary_number}
\end{equation}
To test this model, we plot the experimentally measured $\Fpeak$ rescaled against $\mathrm{Ca}$ in Fig.~\ref{fig:peak_force}. The data collapses onto a master curve bounded by Eqn.~\ref{eqn:rescaled_viscous_force_capillary_number} for the experimentally observed gap limits $\Speak \in [0.1, 1]~\mathrm{mm}$ (see Fig.~S7). We observe a slight deviation in the rescaled data from $\Speak = 1~\mathrm{mm}$, which primarily corresponds to the highest separation velocities. This likely reflects the increasing limitations of shear-rheology-based viscosity estimation under these strongly stretched polymer-bridge conditions. Overall, the collapse shows that, despite complex polymer-solvent interactions that modify the bulk viscosity, at first order the liquid-bridge strength $\Fpeak$ is set mainly by capillary forces and viscous dissipation.

\medskip

\subsection{Role of Particle-Size on Peak Force}


\begin{figure}[htpb]
\centering
  \includegraphics[width=0.49\textwidth]{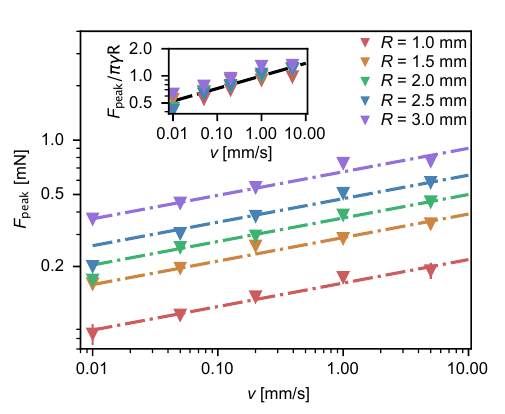}
  \caption{Peak axial force, $\Fpeak$ for a liquid bridge of 1\% 4M PEO solution measured for various particle radius $R \in \left[1, 3\right]\,\mathrm{mm}$. Inset: Rescaled peak force, $\Fpeak/ \pi \gamma R$, which is independent of particle size.}
  \label{fig:force_v_size}
\end{figure}

To quantify the geometric dependence of the axial force, we measure the force for a fixed polymer concentration (1\% 4M PEO) across particle radii $R = 1.0$ to $3.0~\mathrm{mm}$. With increasing particle size, we also scale the bridge volume such that $R \propto V^{1/3}$, resulting in a volume range $V \in [0.12, 3.4]~\mathrm{\mu L}$. The evolution of the measured force as a function of the separation gap is shown in Fig.~S8 of the Supplementary Material, where we note distinct Newtonian and viscoelastic regimes across the various particle sizes. We also observe that the rescaled axial force, $F/\pi\gamma R$, collapses onto a single curve.

\smallskip

The peak axial force for various $R$ is plotted in Fig.~\ref{fig:force_v_size} over a range of separation velocities, increasing with velocity according to a power-law behavior. Because the exponent remains consistent across different particle sizes, we can directly rescale the peak force. As shown in the inset of Fig.~\ref{fig:force_v_size}, normalizing the peak force by the particle size ($F_{\mathrm{peak}} / \pi \gamma R$) successfully collapses the data across the full range of radii. This recovers behavior similar to that of Newtonian liquids, in which capillary-bridge-driven axial forces scale with particle size~\cite{sharma2025experimental}.

\medskip

\subsection{Rupture Distance for the Liquid Bridge}
\label{subsec:rupture_distance}


\begin{figure}[htpb]
\centering
  \includegraphics[width=0.49\textwidth]{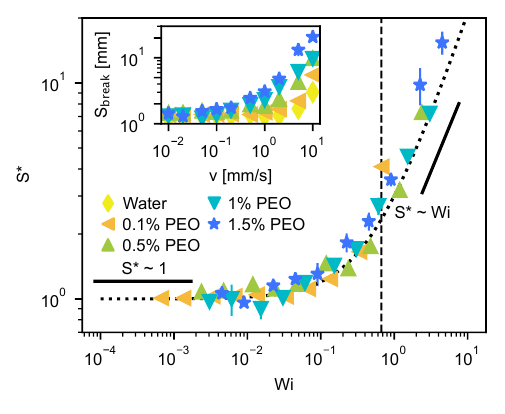}
  \caption{The rescaled rupture distance $S^{*} = \Scv/\Scn$ with respect to the Weissenberg number, $\mathrm{Wi}$, of the polymers. The rescaled $S^{*}$ collapses onto a curve described by Eqn.~\ref{eqn:break_up_distance_rescaled}. Inset shows the experimentally measured rupture distance, $\Srup$. }
  \label{fig:break_up_distance}
\end{figure}

For Newtonian fluids, the rupture distance $\Srup$ is the maximum gap between the particles at which the liquid bridge remains stable~\cite{lian1993theoretical, kruyt_analytical_2017}. However, as noted in previous sections, the viscoelastic regime in polymer solutions modifies it. In these solutions, exponential thinning is typically followed by either satellite drop formation~\cite{Wagner2005} or a blistering instability~\cite{sattler2012final, sattler2008blistering, deblais2018pearling}. The magnitude of the axial force during this instability regime, however, is negligible ($F/\pi \gamma R < 0.01$) and unlikely to influence bulk granular kinematics. Consequently, we adopt this force limit as the threshold for defining the effective rupture distance for the polymeric liquid bridges investigated here.

The approximate rupture distance for a quasi-static Newtonian liquid bridge is a function of bridge volume, $V$, and contact angle, $\theta$ (with $\theta$ expressed in radians), and may be written as~\cite{lian1993theoretical}:
\begin{equation}
    \Srup \approx (1 + 0.5\theta)V^{\frac{1}{3}}
    \label{eqn:rupture_newt}
\end{equation}
For the bridge volumes used in the present work ($V \in [0.5, 1]~\mathrm{\mu L}$), this yields $\Srup \in [1.14, 1.44]~\mathrm{mm}$, which agrees well with experimental observations of $\Srup \in [1.25, 1.89]~\mathrm{mm}$ for quasi-static liquid bridges of 4M PEO ($c = 0$--$1.5\%$) shown in Fig.~\ref{fig:quasistatic_dynamics}(b). We note a slight increase in $\Srup$ with $c$. In this section, we define the Newtonian rupture distance as $\Srup = \Scn$. 

\smallskip

Equation~\ref{eqn:rupture_newt} does not hold for dynamic liquid bridges of polymer solutions, where the rupture distance is significantly larger ($\Scv \gg \Scn$). In Fig.~\ref{fig:break_up_distance}, we plot the non-dimensionalized rupture distance $S^* = \Scv/\Scn$, rescaled with respect to the Weissenberg number imposed on the system by the stretching velocity, $\mathrm{Wi} = \lambda_{\mathrm{R}} \epsv$. Here, any dynamic effects introduced by the polymer are accounted for in the Weissenberg number. Physically, $\Scn$ represents the distance at which a liquid bridge without any polymers, separated quasistatically, would rupture, as described using Eqn.~\ref{eqn:rupture_newt}. For the collapse in Fig.~\ref{fig:break_up_distance}, $\Scn$ is estimated using a representative bridge volume $V = 0.75~\rm \mu L$, corresponding to the approximate average volume across the measurements. This choice likely contributes part of the residual scatter because the individual bridge volumes span $V \in [0.5,1]~\mu\mathrm{L}$. We prefer this quasistatic normalization to a dynamic Newtonian rupture distance, since a reliable model for the rate dependence of Newtonian rupture is not yet available. The relaxation time, $\lambda_{\mathrm{R}}$, is obtained from droplet pinch-off measurements (see Supplementary Material S3). Over the parameter range explored here, rescaling $\Scv$ for polymer solutions collapses the data onto the empirical relation:
\begin{equation}
    S^* = 1 + 2\mathrm{Wi}
    \label{eqn:break_up_distance_rescaled}
\end{equation}
Equation~\ref{eqn:break_up_distance_rescaled} fits the evolution of $S^*(\mathrm{Wi})$ very well. We highlight that rescaling by $\mathrm{Wi}$ rather than $\mathrm{Ca}$ is appropriate here, as the elasto-capillary balance dominates the rupture distance in the viscoelastic regime. At small $\epsv$ ($\mathrm{Wi} \in [10^{-4}, 10^{-2}]$), we find $S^* \sim 1$, aligning with the Newtonian critical rupture gap $\Scn$. At larger imposed rates ($\mathrm{Wi} > 10^{-2}$), the scaling shifts to $S^* \sim \mathrm{Wi}$, confirming that elastic effects control liquid bridge rupture.

\medskip

\section{Conclusions}
Liquid bridges between solid particles are ubiquitous in contexts ranging from soils~\cite{haines1925studies, carminati2017liquid, benard2021physics} to industrial granules~\cite{reynolds2022extensional, lee2024extensional}. In many of these systems, the liquid phase exhibits heterogeneity due to the presence of dissolved polymers~\cite{malarkey2015pervasive}. However, the influence of such heterogeneities on liquid bridge dynamics remains poorly understood, as previous studies have predominantly focused on ideal, fully wetting Newtonian fluids like silicone oil~\cite{willett_capillary_2000, pitois_liquid_2000} or water~\cite{kruyt_analytical_2017, wang_capillary_2017}. In the present work, we investigate polymeric liquid bridges suspended between two solid particles. We quantify the evolution of the axial force as a function of the separation gap across a range of polymer concentrations, separation velocities, and particle sizes.

\smallskip

In the quasi-static regime ($v = 0.01~\mathrm{mm/s}$), we demonstrated---both qualitatively through the shape of the liquid bridge meniscus and quantitatively via force measurements and scaling arguments---that the axial forces are dominated by capillarity. Within the range of polymer concentrations investigated ($2 < c/c^* < 30$), contributions from viscosity and elastic stresses remain negligible. Under dynamic conditions ($v \geq 0.02~\mathrm{mm/s}$), however, we observe and quantify the growing relevance of viscoelastic effects with increasing velocity. To capture this behavior, we divided the axial force evolution into two discrete regimes---Newtonian and viscoelastic---to derive a first-order approximation using parameters obtained solely from shear rheology and thinning dynamics. We then described the evolution of the peak axial force, $\Fpeak$, across various velocities and concentrations through a Capillary-number-based rescaling, and confirmed that $\Fpeak$ scales linearly with the particle radius $R$. Finally, to account for the significantly extended rupture distance in polymer solutions, we introduced a rescaling based on the dimensionless Weissenberg number. This approach allows us to describe the modified rupture distance across a wide range of $v$ and $c$ using a single equation.

\smallskip

This study extends capillary-force measurements to polymeric liquid bridges, beyond earlier work focused on fully wetting liquids~\cite{pitois_liquid_2000, willett_capillary_2000, ennis_1990_influence} or water~\cite{wang_capillary_2017, mielniczuk_characterisation_2018}. We also note a few limitations of the present work. First, our study focused on purely extensional flows between perfectly spherical particles. In real granular systems, however, liquid bridges experience a combination of extension and shear~\cite{song_lateral_2021, song_situ_2025}. While shear contributions can sometimes be approximated using a geometric correction factor~\cite{nase2001discrete}, the shear-thinning nature of polymer solutions introduces additional complexity that warrants further investigation. Additionally, real granular media consist of irregularly shaped particles, which present another necessary avenue for follow-up work~\cite{butt_normal_2009, chatterjee2013effect}.

\smallskip

Despite these limitations, the results provide a practical first-order framework for modeling bulk cohesive granular flows, such as in Discrete Element Method (DEM) simulations~\cite{guo_discrete_2015, radjai2017modeling}. When the bridge volume and contact angle are known accurately, more detailed closed-form or approximate capillary-force expressions are also available for efficient DEM implementations in the quasi-static limit~\cite{argilaga_closedform_2023,bagheri_approximate_2024}. By dividing the axial force into distinct Newtonian and viscoelastic regimes, the attractive normal force in a DEM implementation may be written as
\begin{equation}
    F_n = \begin{cases}
    -\pi \gamma R_{\mathrm{min}} - F_{\mathrm{visc}}, & \text{for } 0 \le \delta_n \le S_{\mathrm{c}}, \\
    -F_{\mathrm{c}} e^{-\frac{\delta_n - S_{\mathrm{c}}}{3v\lambda_{\mathrm{e}}}}, & \text{for } S_{\mathrm{c}} < \delta_n \le S_{\mathrm{rupt, elast}}, \\
    0, & \text{for } \delta_n > S_{\mathrm{rupt, elast}},
    \end{cases}
\end{equation}
where the negative sign denotes attraction and $S_{\mathrm{rupt, elast}}$ is defined by Eqn.~\ref{eqn:break_up_distance_rescaled}. Such reduced particle-scale laws are particularly useful for connecting bridge-scale physics to macroscopic behaviors observed in cohesive granular systems, including, for instance, wet accretion and cohesive granular flows~\cite{saingier2017accretion, gans2023collapse, sharma2024effects}.

\smallskip

Future work could refine this model by developing a continuous, rather than piecewise, description of the transition from the Newtonian to the viscoelastic regime, with a specific focus on the critical force $F_{\mathrm{c}}$. While Eqn.~\ref{eqn:Critical_force_transition} offers a preliminary model for this boundary, a rigorous study of the transition dynamics is needed. Ultimately, by capturing the influence of polymer additives on bridge rupture and cohesive strength, this framework takes a first step towards bridging the gap between micro-scale rheology and the macroscopic mechanics of polymer-based cohesive granular materials.

\medskip

\section*{Conflicts of interest}
There are no conflicts to declare.

\section*{Acknowledgments}

This work was supported by the National Science Foundation under NSF CAREER Program Award CBET Grant No. 1944844, NSF PMP Grant No. 2533460, by the U.S. Army Research Office under Grant No. W911NF-23-2-0046, by the International Fine Particle Research Institute, and by the Gordon and Betty Moore Foundation, Grant DOI 10.37807/GBMF13831. The authors thank Anuj Acharya for their help in preliminary studies.

\clearpage
\bibliographystyle{unsrtnat}
\bibliography{Axial_forces_arxiv}

\clearpage


\section*{Supplementary material}

# Axial forces in capillary liquid bridges of polymer solutions

Sreeram Rajesh[1], Riley S. Tinianov[1], Jooyeon Park[2], and Alban Sauret[2,3]

[1] Department of Mechanical Engineering, University of California, Santa Barbara, California 93106, USA
[2] Department of Mechanical Engineering, University of Maryland, College Park, Maryland 20742, USA
[3] Department of Chemical and Biomolecular Engineering, University of Maryland, College Park, Maryland 20742, USA

## S1  Validating the experimental setup with silicone oil

To validate our force sensor setup, we measure axial force in a capillary liquid bridge formed by AP100 silicone oil (by Sigma-Aldrich). AP100 has very low viscosity ($\eta \simeq 100$ mPa.s) and surface tension ($\gamma = 24$ mN/m, as provided by the supplier), which also results in a small contact angle ($\theta = 0°$). For the validation test, the volume of silicone oil is also kept low to avoid any perturbation from gravity. We measure the volume $V = 0.036$ $\mu$L from the image analysis. The particles are separated quasi-statically. We show the capillary force ($F$) measured as the gap ($S$) increases in Fig. S1. The measured force is compared with the approximation derived from the Gorge method by Pitois *et al.* [1]:

$$F_{\text{Cap}} \approx 2\pi R \gamma \cos\theta \left[ 1 - \frac{1}{\sqrt{1 + \frac{2V}{\pi R S^2}}} \right] \tag{S1}$$

The experimental measurements agree well with this approximate equation, which is only valid for liquids with small contact angle ($\theta \simeq 0-10°$). Notably, for $V = 0.036$ $\mu$L, the experimental measurements follow the expected curve very closely. Furthermore, the measured rupture distance $S_{\text{rup}} = 0.345$ mm is close to the value estimated from the expression provided by Lian *et al.* [2] $(1 + 0.5\theta)V^{1/3} \simeq 0.33$ mm.

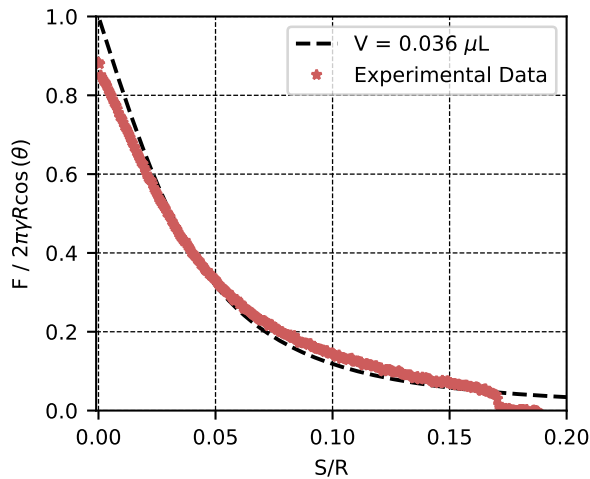

Figure S1: Axial force due to AP100 silicone oil of $V = 0.036$ $\mu$L on spherical particles measured using the force sensor based setup described in the main article.

# S2 Shear Rheology: Viscosity and Modulus for 4M PEO solutions

We measured the shear viscosity ($\eta$), storage modulus ($G'$), and loss modulus ($G''$) of 4M PEO solutions ($M_\mathrm{w}$ = 4000 kg/mol) dissolved in water using a rheometer (Anton Paar MCR 302) with a 50 mm, 1° cone-plate geometry. The viscosity increases with concentration, and clear shear-thinning is observed for the polymer solutions shown in Fig. S2(a). The corresponding storage and loss moduli are reported in Fig. S2(b).

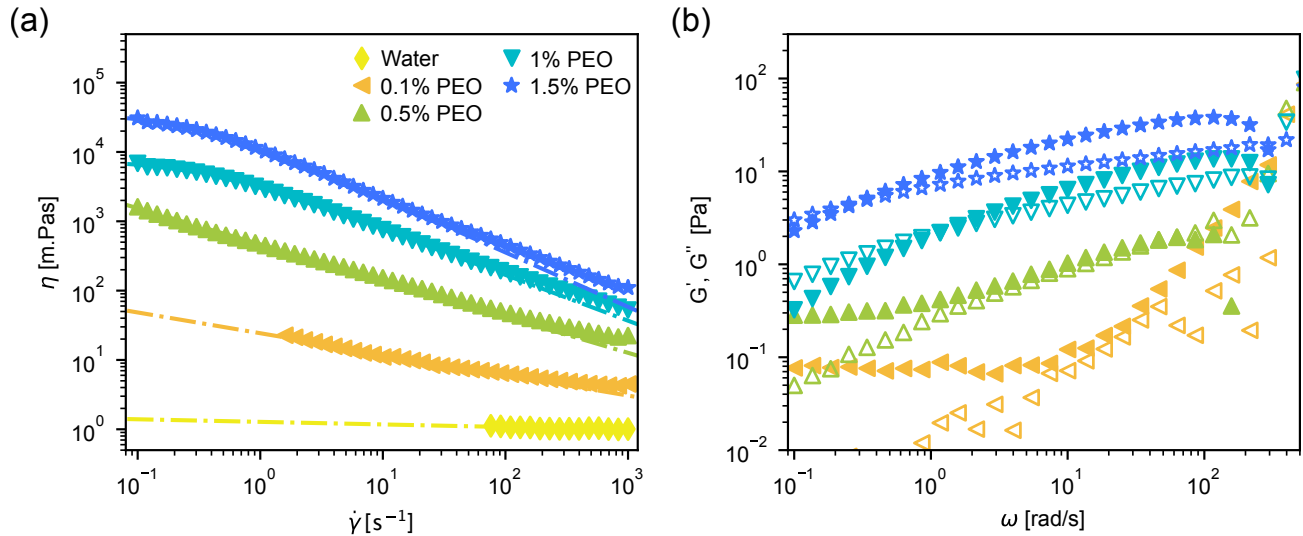

Figure S2: (a) Evolution of the shear viscosity $\eta$ with shear rate $\dot\gamma$. (b) Storage modulus $G'$ (solid symbols) and loss modulus $G''$ (open symbols) as functions of angular frequency $\omega$ for water and 4M PEO solutions at different concentrations.

# S3 Extensional Rheology: Relaxation Time for 4M PEO solutions using Pendant Drop Method

To measure the relaxation time, we quantify the evolution of the minimum thickness of the neck in a droplet pinch-off, $h_{\min}(t)$. As noted in numerous previous studies [3, 4], the droplet pinch-off in the viscoelastic regime follows an exponential thinning characterized as: $h_{\min}(t) \propto e^{-t/3\lambda_R}$. By fitting the thinning dynamics in the exponential regime, we extract the relaxation time ($\lambda_R$) for the polymer solutions, summarized in Fig. S3. As noted in previous works, the relaxation time scales with the concentration of the polymer solution as

$$\lambda_R \propto c^{0.7} \tag{S2}$$

where the exponent of 0.7 is similar to values previously reported by Tirtaatmadja *et al.* [4] and Rajesh *et al.* [3].

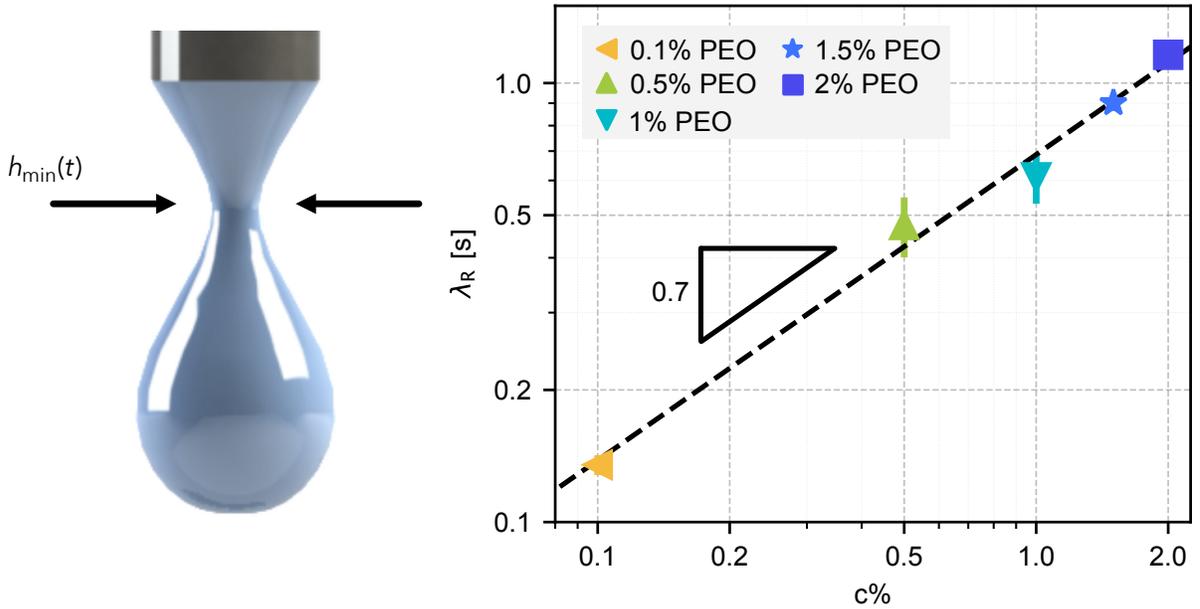

Figure S3: (a) Schematic of a droplet indicating how the minimum thickness of the thinning droplet is measured (b) Relaxation time of a 4M PEO solution at $c = 0.1 - 2$ % in water. The nozzle diameter used is $h_0 = 2.75$ mm.

3

# S4 Surface Tension and Dynamic Solid-Liquid Contact Angle

In Fig. S4(a), we show the surface tension ($\gamma$) for the 4M PEO polymer solutions of various concentrations ($c$) prepared in water. As $c$ increases, $\gamma$ indeed decreases. For $c = 0.1 - 1\%$, we are able to systematically quantify this using a tensiometer (Attension). For $c = 1.5\%$, the tensiometer measurements are inconsistent. However, from $\gamma$ measurements for higher concentrations of high $M_\text{w}$ polymer solutions, similar to the $M_\text{w} = 4000$kg/mol used in this study, noted in a previous work by Cao *et al.* [5], we do not expect $\gamma$ to drop significantly for the $c = 1.5\%$ 4M PEO solution. Hence, we consider $\gamma = 60$ mN/m for the present work.

To quantify the solid-liquid contact angle $\theta$ (see Fig. 1(c) in the main article), we developed a custom image-analysis protocol to measure $\theta$ as a function of the separation gap $S$. In Fig. S4(b), we show $\theta(S)$ under quasi-static conditions ($v = 0.01$ mm/s). As shown here, $\theta$ decreases slightly with increasing $c$, which is expected due to the decrease in $\gamma$. This trend is slightly off for $c = 1\%$, but as noted in the main article, this deviation is attributed to the slightly lower volume of liquid dispensed in the liquid bridge.

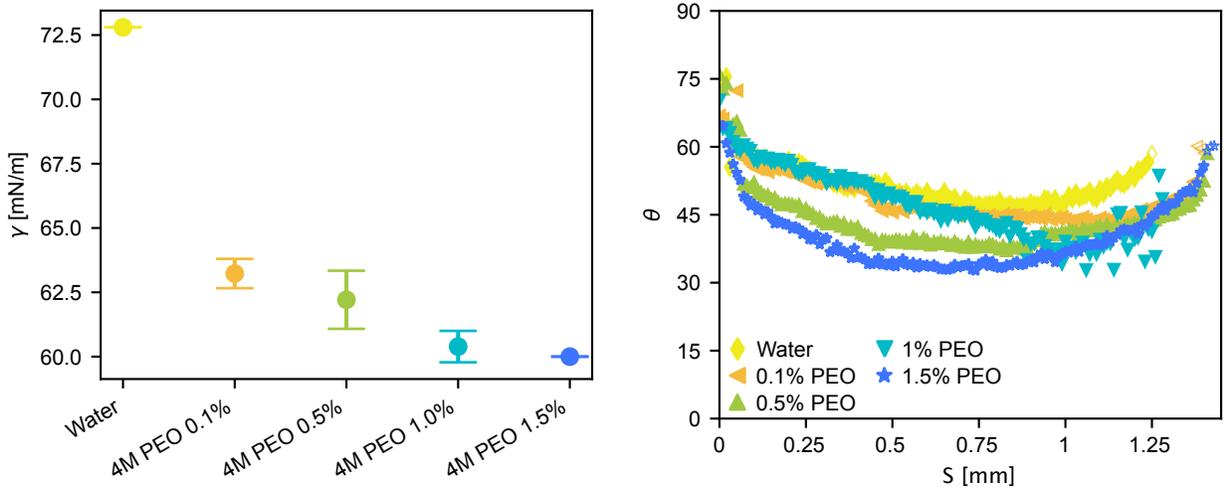

Figure S4: (a) Surface tension and (b) evolution of the contact angle ($\theta$) as a function of the gap ($S$) between the particles for water and 4M PEO solutions of $c = 0.1 - 1.5$ %.

# S5   Volume Measurement for the Liquid Bridge

The volume of the liquid bridge was calculated by treating the binarized 2-dimensional profile (see Fig. S5) as a solid of revolution. Assuming axisymmetry about the vertical axis, the total volume $V$ is defined by the integral:

$$V = \int_{z_{min}}^{z_{max}} A(z)\,dz = \int_{z_{min}}^{z_{max}} \frac{\pi D(z)^2}{4}\,dz \tag{S3}$$

where $D(z)$ is the local diameter of the bridge at height $z$. To compute this from the image, the domain was discretized into $N$ horizontal slices corresponding to the pixel rows. The volume was numerically approximated using a sum:

$$V \approx \sum_{i=1}^{N} \frac{\pi d_i^2}{4} \Delta z \tag{S4}$$

In this expression, $d_i$ is the physical diameter of the bridge in row $i$, obtained from the pixel count $n_i$ through $d_i = n_i/\alpha$, where $\alpha$ is the image calibration factor in pixels per mm. Likewise, $\Delta z = 1/\alpha$ is the physical height of a single pixel row. This summation yields the total volume in mm$^3$ (equivalent to $\mu$L).

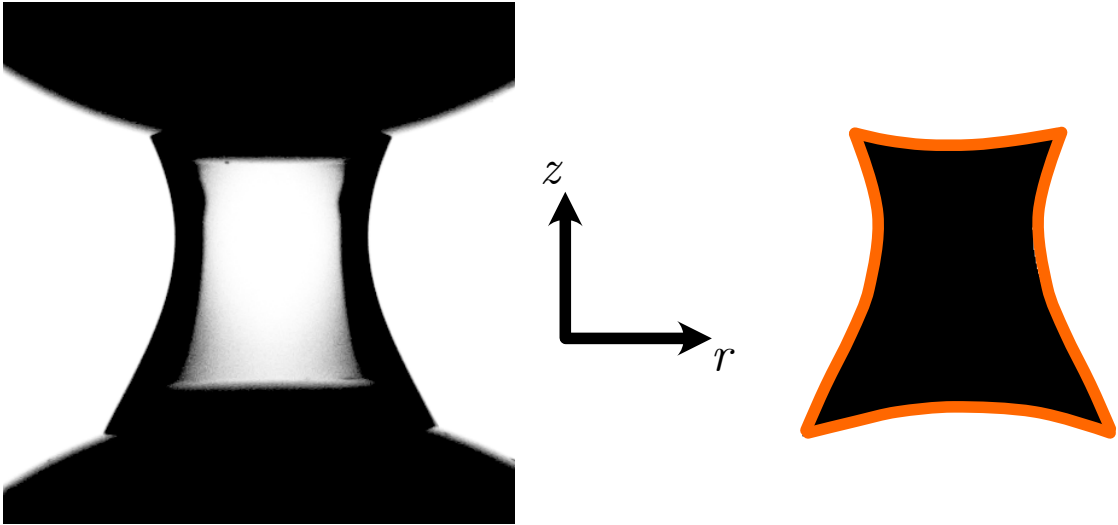

Figure S5: Contour used for measuring liquid bridge volume.

5

# S6 Elastic Prefactor and Relaxation Time for $v = 0.01 - 10$ mm/s

In Fig. 5 in the main article, we fit the axial forces in the viscoelastic regime with the assumption that the force decays exponentially, where the resulting expression has the form:

$$F = F_\text{c} \exp\left(-\frac{S - S_\text{c}}{3v\lambda_\text{e}}\right) \tag{S5}$$

Here, $S$ is the gap between the particles, $S_\text{c}$ is the critical gap at the onset of the viscoelastic regime, and $\lambda_\text{e}$ is the characteristic thinning timescale estimated from the recorded ligament dynamics. The only fitting parameter is $F_\text{c}$, the transition-force prefactor, which is summarized in Fig. S6(a) for 4M PEO at $c = 1\%$. We observe that $F_\text{c}$ approaches a nearly constant value as $v$ increases, while $\lambda_\text{e}$ remains approximately independent of $v$.

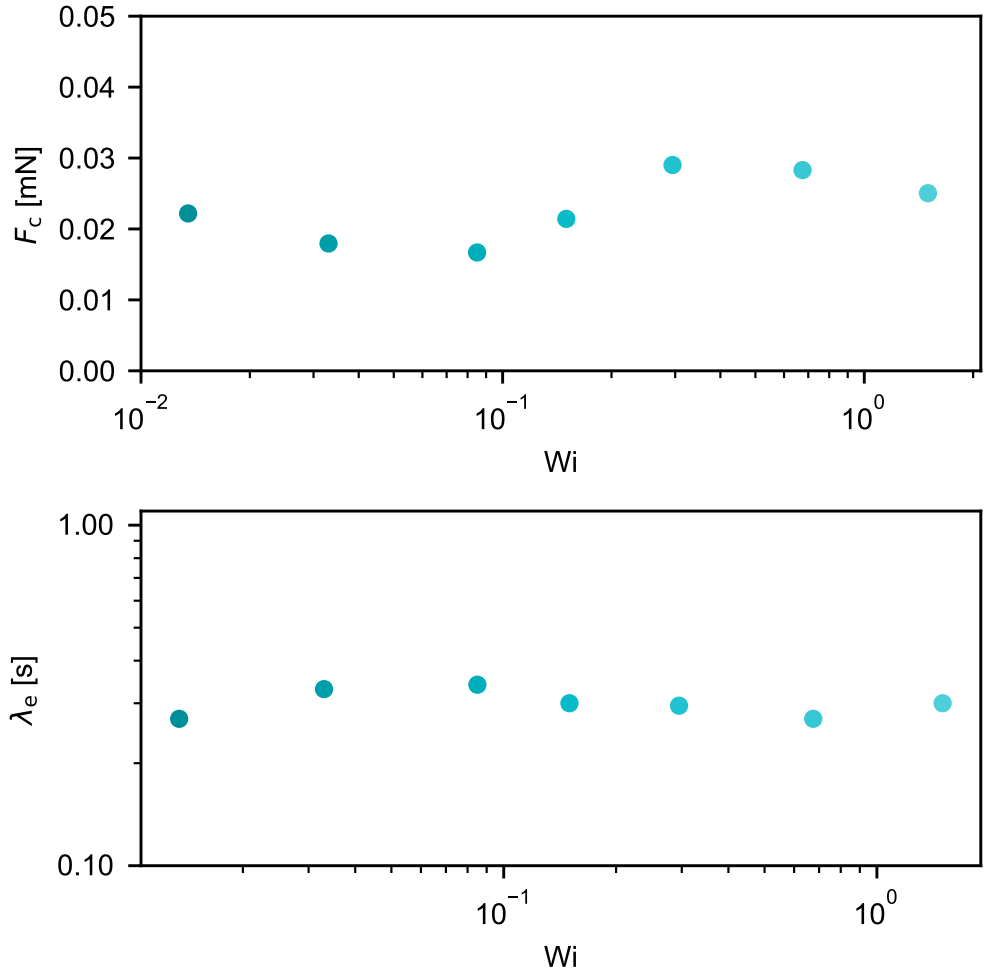

Figure S6: (a) Transition force $F_\text{c}$ extracted from the exponential fit in the viscoelastic regime. (b) Characteristic timescale $\lambda_\text{e}$ measured from the thinning dynamics of the liquid bridge for various separation velocities $v$.

# S7 Particle-Particle Gap $S_{\text{peak}}$ for Peak Axial Force

We define the gap, $S$ between the particles where we measure the peak axial force during the particle separation (shown in Fig. 6 in the main article) as $S_{\text{peak}}$. In Fig. S7, we show $S_{\text{peak}}$ for different concentrations of the polymer solutions for the range of velocities investigated in this work ($v = 0.01 - 10$ mm/s). The range considered for $S_{\text{peak}} \in [0.1 - 1]$ mm for the theoretical fits, shown in Fig. 6, is indeed reasonable based on experimental observations.

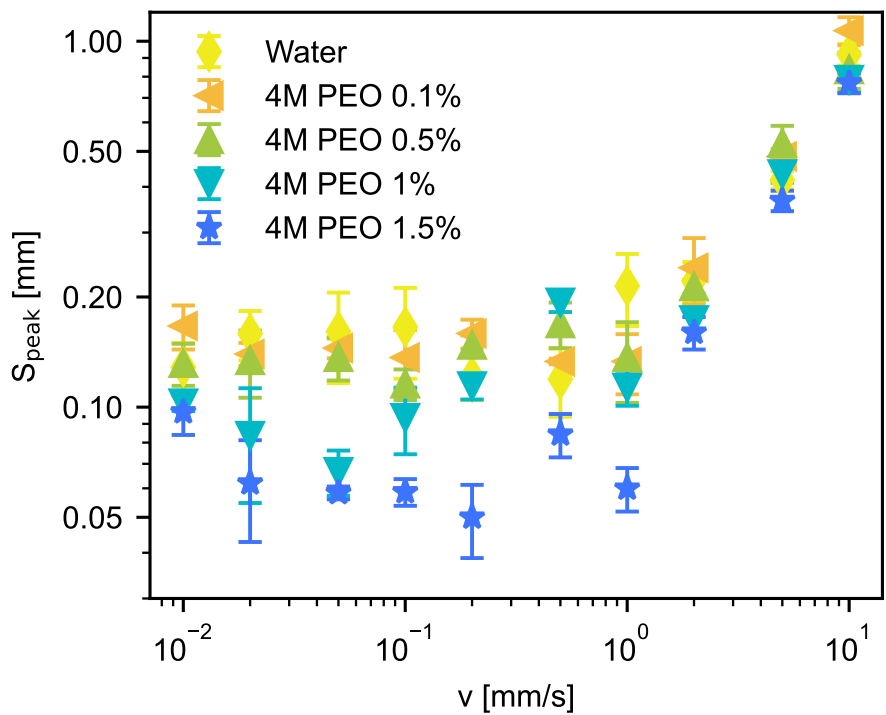

Figure S7: Gap corresponding to the peak force, $S_{\text{peak}}$, recorded for solutions of various concentrations at $v = 0.01 - 10$ mm/s.

## S8  Role of Particle-size

Figure S8(a) shows the axial force as the particle separation increases at a velocity $v = 5$ mm/s for particle radii $R \in [1, 3]$ mm. The corresponding liquid bridge volumes are $V = 0.126, 0.42, 1, 2,$ and $3.4$ $\mu$L, respectively, chosen to approximately maintain the scaling $R \propto V^{1/3}$. The axial force exhibits initial Newtonian decay followed by a transition to the viscoelastic regime, with the final rupture distance depending heavily on $V$.

Figure S8(b) presents the evolution of the rescaled force, $F/\pi\gamma R$. The rescaled curves largely collapse across the different particle sizes. The slight deviation observed for $R = 1$ mm arises from experimental limitations in accurately loading such a small sample volume. Overall, these results illustrate that the standard scaling of capillary force with particle size remains valid for polymeric liquid bridges.

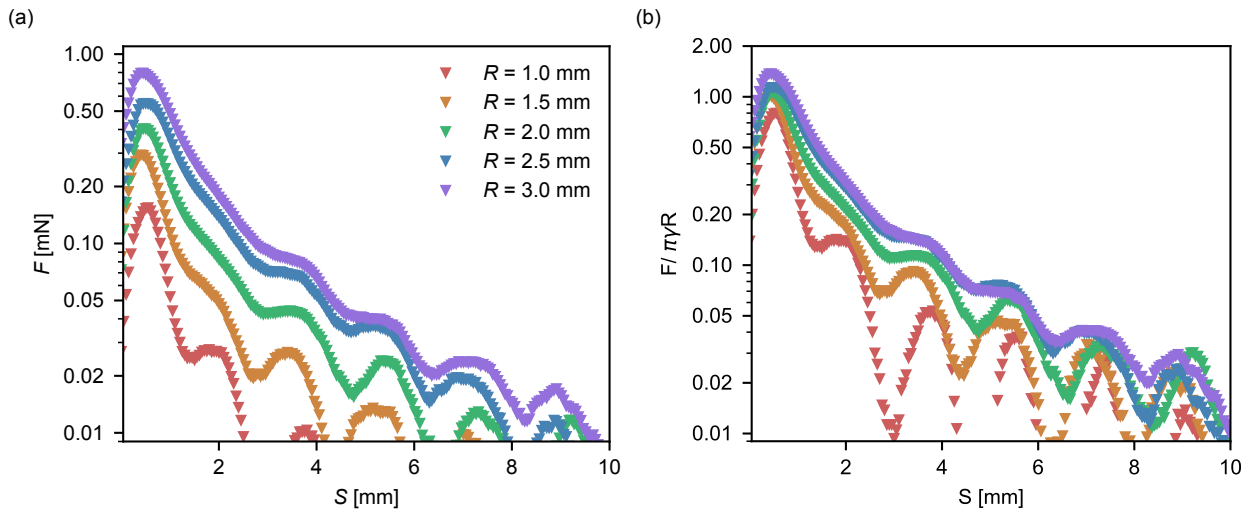

Figure S8: (a) Evolution of the axial force $F$ with separation distance $S$ for a 1 wt% 4M PEO solution at a fixed velocity $v = 5$ mm/s across varying particle radii $R$. (b) The rescaled evolution $F(S)/\pi\gamma R$.



\end{document}